\documentclass[pra,twocolumn]{revtex4}
\usepackage{graphicx,amsmath,amssymb,float,bm,times}

\newcommand{\ek}{\epsilon_{\mathbf{k}}}

\newcommand{\Ek}{E_{\mathbf{k}}}

\newcommand{\Omegaq}{\Omega_{\mathbf{q}}}

\newcommand{\uk}{u_{\mathbf{k}}}
\newcommand{\vk}{v_{\mathbf{k}}}
\newcommand{\createa}[1]{a^\dagger_{#1}}
\newcommand{\destroya}[1]{a^{\phantom \dagger}_{#1}}
\newcommand{\createb}[1]{b^\dagger_{#1}}
\newcommand{\destroyb}[1]{b^{\phantom\dagger}_{#1}}

\begin{document}

\title{Particle density distributions in Fermi gas superfluids:
  Differences between one and two channel models}

\author{Jelena Stajic$^1$, Qijin Chen$^2$, and K.  Levin$^1$}

\affiliation{$^1$ James Franck Institute and Department of Physics,
  University of Chicago, Chicago, Illinois 60637}

\affiliation{$^2$ Department of Physics and Astronomy, Johns Hopkins 
University, Baltimore, Maryland 21218}

\begin{abstract}
  We show how to describe the $T \neq 0$ behavior associated with the
  usual BCS- Bose Einstein condensation (BEC) crossover ground state.
  We confine our attention here to the BEC and near-BEC regime where
  analytical calculations are possible.  At finite $T$, non-condensed
  fermion pairs must be included, although they have been generally
  ignored in the literature.  Within this BEC regime we compute the
  equations of state for the one and two channel models; these two cases
  correspond to whether Feshbach resonance effects are omitted or
  included.  Differences between these two cases can be traced to
  differences between the nature of a Cooper pair and bosonic
  condensate.  Our results are also compared with the Gross Pitaevskii
  equations of state for true bosons. Differences found here are
  associated with the underlying fermionic character of the system.
  Finally, the particle density distribution functions for a trap
  containing superfluid fermionic atoms are computed using a
  Thomas-Fermi approach. The one and two channel behavior is found to be
  very different; we find a narrowing of the density profile as a result
  of Feshbach resonance effects. Importantly, we infer that the ratio
  between bosonic and fermionic scattering lengths depends 
  on the magnetic detuning and is generally smaller than 2.
  Future experiments will be required to determine to what extent this
  ratio varies with magnetic fields. 
\end{abstract}
\maketitle

\section{Introduction}
\label{sec:1}

The recent observations
\cite{Jin3,Grimm,Ketterle2,Salomon3,Jin4,Ketterle3,Thomas2,Grimm3} 
of Bose-Einstein
condensation (BEC) of molecules formed from fermionic atoms are
extremely exciting.
%Moreover, some would argue that
%these experiments, 
%which probe the crossover from the BCS to the Bose-Einstein
%regime, have implications for high $T_c$ superconductivity. These
%measurements will, at the least, provide the community with a
%deeper understanding of super-conductivity and fluidity
%as the attraction between fermions varies from weak to strong.
%
Because of a Feshbach resonance, in which fermions couple to molecular
bosons, it is possible  \cite{Holland,Timmermans}, via application of
magnetic fields, to obtain an arbitrarily strong attraction between
fermions, and to probe the crossover \cite{Griffin,Milstein} from BEC to
BCS.  The resultant superfluidity of pre-formed pairs has a natural
counterpart in some theories  \cite{Ranninger2,Micnas,Chen2,Chen4} of high
$T_c$ superconductors. Indeed, a striking, and important feature of the
cuprate superconductors is their pronounced precursor superconductivity,
or ``pseudogap" effects, the origin of which is still under active
debate.
%MAYBE WE NEED TO DISCUSS SCATTERING LENGTHS-- HOW THERE ARE 3 REGIMES:
%BCS, BEC AND RESONANCE REGIME.

A second, very important motivation for these experiments is based on
the theoretical observation that a BCS-like ground state wavefunction
is capable \cite{Leggett} of describing both fermionic and
bosonic-based superconductors, provided that the chemical potential of the
fermions, $\mu$, is determined self consistently.  Given the vast
success of weak coupling or BCS theory, it is extremely important to
formulate this extended theory at all $T$, and confront it with
controlled experiments.  

In this paper we explore the implications of this specific ground state
wavefunction \cite{Eagles,Leggett} in detail to address both $T=0$ and
$T=T_c$, with particular emphasis on the ``near-BEC" regime.  Our goals
are (i) to emphasize how to include finite temperature effects in a
manner consistent with the ground state; this necessitates the
introduction of non-condensed fermion pairs which lead to ``pseudogap"
effects \cite{JS2}, (ii) to provide analytic calculations and insights by
working in a (near-BEC) regime where calculations are more tractable,
(iii) to discuss in some detail the differences between the ``one
channel" and ``two channel" models for the Feshbach resonance, and (iv)
to compare with well established theories of weakly interacting Bose
superfluid  \cite{RMP}, as well as (v) with the measured density
distributions in a trap.  We begin with the homogeneous case and then
consider the trap configuration; from this we infer the ratio between
the effective bosonic and fermionic scattering lengths, which is found
to be strongly dependent on the magnetic detuning $\nu_0$, and generally
less than 2, as a consequence of many-body combined with Feshbach
resonance effects.

\section{T matrix Formalism: Beyond Nozieres and Schmitt-Rink}
\label{sec:2}

The Hamiltonian used in the cold atom and high $T_c$
crossover studies consists,
in its most general form,
of two types of interaction effects: those associated with the direct
interaction between fermions parameterized by $U$, and those associated
with ``fermion-boson" interactions, whose strength is governed by $g$.

\begin{eqnarray}
\label{hamiltonian}
H&-&\mu N=\sum_{{\bf k},\sigma}(\epsilon_{\bf k}-\mu)\createa{{\bf
 k},\sigma}\destroya{{\bf k},\sigma}+\sum_{\bf q}(\epsilon_{\bf
 q}^{mb}+\nu-2 \mu)\createb{\bf q}\destroyb{\bf q}\nonumber\\
&+&\sum_{{\bf q},{\bf k},{\bf k'}}U({\bf k},{\bf k'})\createa{{\bf
 q}/2+{\bf k},\uparrow}\createa{{\bf q}/2-{\bf
 k},\downarrow}\destroya{{\bf q}/2-{\bf k'},\downarrow}\destroya{{\bf
 q}/2+{\bf k'},\uparrow}\nonumber\\
 &+&\sum_{{\bf q},{\bf k}}\left(g({\bf k})\createb{{\bf q}}\destroya{{\bf
  q}/2-{\bf k},\downarrow}\destroya{{\bf q}/2+{\bf
 k},\uparrow}+h.c.\right)
 \label{eq:0c}
 \end{eqnarray}
%  $U({\bf k},{\bf  k'})=U\varphi_{\bf k} \varphi_{\bf k'}$,
 %$g({\bf k})=g\varphi_{\bf k}$, $\varphi_{\bf k}^2 = \exp\{-(k/K_c)^2\}$
 % \hskip 1cm
 Here the fermion and boson kinetic energies are given by $\ek\equiv
 \hbar^2 k^2/2 m$ and $\epsilon_{\bf q}^{mb} \equiv \hbar^2 q^2/2 M$,
 respectively, and $\nu$ is an important parameter which represents the
 ``detuning".
The variational ground state which we will consider here is
a product of both fermionic and bosonic contributions
\begin{equation}
 \bar{\Psi}_0 = \Psi_0 \otimes \Psi_0^B
\end{equation}
where the normalized fermionic wave function is the standard 
crossover state \cite{Eagles,Leggett} 
\begin{equation}
\Psi_0=\Pi_{\bf k}(\uk+\vk c_k^{\dagger} c_{-k}^{\dagger})|0\rangle
\label{eq:1a}
\end{equation}
and the normalized molecular or Feshbach boson contribution $\Psi_0^B$
is represented by
\begin{equation}
 \Psi_0^B=
e^{-\lambda^2/2+\lambda b_0^{\dagger}}|0\rangle \,.
\end{equation}
The variational parameters are, thus, $\uk$, $\vk$ and $\lambda$.
Applying standard variational techniques on the ground state
wavefunction leads to 
Eqs.~(\ref{eq:extra2}) and (\ref{eq:extra}), 
and the $T=0$ limit of Eq.~(\ref{number_equation}).
%the $T=0$ limits of
%Eqs.~(\ref{eq:gap_equation}) and (\ref{eq:number_equation}) below, along
along with the result that $\lambda = \phi _m \equiv\langle b_{{\bf q}=0}
\rangle$.

Whether both forms of interactions ($U$ and $g$) in the Hamiltonian are
needed in either system is still under debate. The bosons ($\createb{\bf
  k}$) of the cold atom problem  \cite{Holland,Timmermans} will be
referred to as Feshbach bosons.  These represent a separate species, not
to be confused with the fermion pair ($\createa{\bf k} \createa{\bf
  -k}$) operators.  Thus we call this a ``two channel" model.
%It is also possible
%to contemplate systems where there are two types of fermions,
%one pair of which can be associated with the Feshbach boson. This
%approach reduces effectively to a one channel model, primarily because
%of the way in which the condensate is handled. (See Section \ref{sec:2D}).

For the sake of clarity we begin by ignoring the Feshbach
resonance-induced interactions.
The variationally induced constraints on the
fermionic degrees of freedom \cite{Leggett}
are given by 
\begin{equation}
0=1+U \sum_{\bf
  k}\frac{1}{2 \Ek}
\label{eq:extra2}
\end{equation}
with the fermion density $n = 2\sum _{\bf k} \vk^2$, where $\vk^2 \equiv
\left[ 1 -(\ek - \mu)/\Ek \right]/2 $. 
We define the quasiparticle dispersion, $\Ek$, below.

One can anticipate that the excited states of this system are associated
with their natural extension  \cite{Chen2,Chen4,JS2} following BCS theory.
For $ T \leq T_c$, the gap equation and number equations now involve
Fermi functions $f(\Ek)$ and are given by
\begin{eqnarray}
\frac{m}{4 \pi \hbar^2 a_s }& =&  \mathop{\sum_{\bf k}}\left[ \frac{1}{2
  \ek } - \frac{1- 2 f(\Ek)}{2E_{\bf k}} \right]\,, \label{eq:gap_equation}\\ 
n & =& \sum _{\bf k} \left[ 1 -\frac{\ek - \mu}{\Ek} 
+2\frac{\ek - \mu}{\Ek}f(\Ek)  \right] \,,
\label{eq:number_equation}
\end{eqnarray}
where 
$\mu$ is now evaluated at finite
$T$ and $\Ek = \sqrt{ (\ek -\mu)^2 + \Delta^2 }$.  
 Here we assume a contact
 potential where $a_s$ is the isolated two body
 scattering length, which is simply related to $U$,
(see, for example, Eq.~(\ref{eq:A6})).

Two important observations need to be made. Here $\Delta$ 
is associated with the
excitation gap for fermions.
\textit{This excitation gap reflects both
condensed and non-condensed pairs and is not
the same as the order parameter $\Delta_{sc}$, except at $T=0$}.  
Eqs.~(\ref{eq:gap_equation}) and (\ref{eq:number_equation}) are
widely cited in the literature, but with an important difference.
Throughout the literature $\Delta$ is taken to be the order
parameter.
By contrast, to determine $\Delta_{sc}$ here we will first need to 
characterize the contribution of non-condensed pairs to
$\Delta$.

Our second important observation is that for
$ T \leq T_c$, the Fermi
factors in the above equations are essentially negligible \cite{Words3}. 
It follows that
\textit{both $\Delta(T)$ and $\mu(T)$
are temperature independent in this near-BEC regime.}  Indeed, this is
consistent with the physical picture of well established,
pre-formed pairs in the BEC limit, so that the fermionic
energy scales are unaffected by $T$ below $T_c$.

\begin{figure}
\includegraphics[width=3.4in,clip]{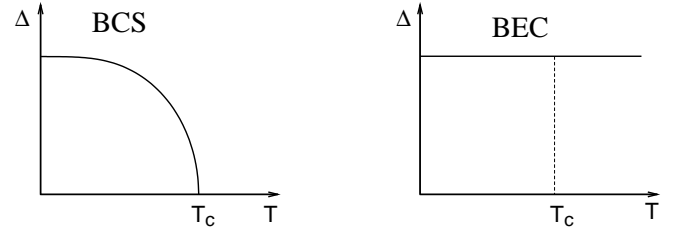}
\caption{Contrasting temperature dependences of $\Delta$ 
in the BCS and BEC regimes. Similarly, in the BEC regime
$\mu$ is a constant, so that all fermionic energy scales are $T$ independent,
as expected.}
\label{fig:3}
\end{figure}

The simple physics in Eq.~(\ref{eq:gap_equation}) may
be schematically represented by plots of $\Delta$ versus
temperature. Figure \ref{fig:3} contrasts the behavior in the 
weak coupling or BCS and strong coupling or BEC
regimes. 
In the BCS limit $\Delta(T)$ follows the behavior of
the order parameter, whereas in the BEC regime, pairs
are pre-formed and there is no temperature dependence in $\Delta(T)$ on the
scale of $T_c$. We now extend these qualitative observations
to a more quantitative level.

\subsection{Extending Conventional Crossover Theory to $T \neq 0$:
BEC Limit Without Feshbach Bosons}
\label{sec:2A}

The self consistent equations in the BEC limit for
general temperature $T$ can then be
written as

\begin{eqnarray}
\frac{m}{4 \pi \hbar^2 a_s } & = & \mathop{\sum_{\bf k}}\left[ \frac{1}{2
 \ek } - \frac{1}{2E_{\bf k}} \right]  \,, \label{eq:3a}\\
 n & =& \sum _{\bf k} \left[ 1 -\frac{\ek - \mu}{\Ek}
  \right] \,, ~~~~T \leq T_c\,.
 \label{eq:4a}
 \end{eqnarray}
%
%Here we have set all Fermi functions to zero (their arguments are of the
%order of $|\mu|/T_c>>1$) in Eqs.~(\ref{eq:gap_equation}) and
%(\ref{eq:number_equation}).
It follows from the above equations that, just
as in the $T=0$ limit \cite{Strinati,Stringari}, we have for
general $ T \le T_c$,
 \begin{equation}
 n_{pairs} =\frac{n}{2}= Z_0 \Delta^2
 \label{eq:11}
 \end{equation}
where the coefficient of proportionality  
\begin{equation}
Z_0 \approx \frac {m^2 a_s } { 8 \pi \hbar^4} \,.
\label{eq:10}
\end{equation}

We arrive at an important physical interpretation.  Even though
$\Delta$ or $n_{pairs}$ is a constant in $T$, this constant must
be the sum of two temperature dependent terms. Just as in the
usual theory of BEC these two contributions correspond to
condensed and non-condensed components 
\begin{equation}
n_{pairs} = n_{pairs}^{condensed}(T) + n_{pairs} ^ {non-condensed}(T) \,,
 \label{eq:12}
 \end{equation}
so that we may decompose the excitation gap into two contributions
 \begin{equation}
 \Delta^2  = \Delta_{sc}^2 (T) + \Delta_{pg}^2 (T) \,,
 \label{eq:13}
 \end{equation}
 where $\Delta_{sc}(T)$ corresponds to condensed and $\Delta_{pg}(T)$ to
 the non-condensed (or pseudo) gap component. Each of these are proportional to the
 respective number of condensed and non-condensed pairs with
 proportionality constant $Z_0$.  Just as in BEC, at $T_c$,
 \begin{equation}
 n_{pairs}^{non-condensed}(T_c) = \frac{n}{2}= \sum_{\bf q} b
 (\Omega_q,T_c) \,,
 \label{eq:52}
 \end{equation}
 where $b(x)$ is the usual Bose-Einstein function and $\Omega_q$
 is the dispersion of the non-condensed pairs, which will
 be self consistently determined below.
 Thus
 \begin{equation}
 \Delta^2(T_c)= \Delta_{pg}^2 (T_c) = Z_0^{-1} \sum b(\Omega_q, T_c) =
 \frac{n}{2} Z_0^{-1} \,.
 \label{eq:14}
 \end{equation}
 We may deduce directly from Eq.~(\ref{eq:14}) that $\Delta_{pg}^2 =
 -\sum_Q t(Q)$, if we presume that below $T_c$, the non-condensed pairs
 have propagator
 \begin{equation}
t(Q) = \frac{Z_0^{-1}}{\Omega- \Omega_q} \,.
\label{eq:15}
\end{equation}

This lead to a key question: how can one deduce the contribution from
\textit{non-condensed} pairs?  We now work backwards to infer the
dispersion $\Omega_q$ for these pairs.  A fundamental requirement on
non-condensed pairs in equilibrium with a Bose condensate is that their
effective chemical potential satisfy
\begin{equation}
 \mu_{pair}(T) = 0,~~~~T \leq T_c \,. 
\label{eq:chempot} 
\end{equation}
Equation (\ref{eq:gap_equation}) 
can be shown to be consistent with
Eq.~(\ref{eq:chempot}) provided that the 
propagator for non-condensed pairs is given by
\begin{equation}
t(Q)= \frac{U}{1+U  \chi(Q)},
\label{eq:t-matrix_pg}
\end{equation}
where
\begin{equation}
\chi(Q) \equiv  \mathop{\sum_K} G(K)G_0(Q-K),
\label{eq:chi}
\end{equation}
%\Sigma_{pg}(K) &=&  \mathop{\sum_Q} t_{pg}(Q)G_0(Q-K).
%
and $G$ represents the fermionic Green's function which has a self
energy $\Sigma(K) = - \Delta^2 G_0 (-K) $. Here $G_0$ is the bare
propagator. The details of this analysis are presented in
Appendix \ref{app:1}.

Another important point should be noted.  This pair propagator
or T-matrix differs from that first introduced by Nozieres and
Schmitt-Rink \cite{NSR} because here there is one dressed and
one bare Green's function. In the approach of Ref.~\onlinecite{NSR}
both are taken as bare Green's functions. By contrast there
are other schemes in the literature \cite{Haussmann,Tchern,YY} where both
Green's functions are dressed.
We end by noting that at small four-vector $Q$ (and moderately strong coupling)  we may
expand Eq.~(\ref{eq:t-matrix_pg}) to obtain
\begin{equation}
t(Q) = \frac { Z_0^{-1}}{\Omega - \Omega_q +\mu_{pair} + i \Gamma_Q}.
\label{eq:expandt}
\end{equation}
Now we can deduce directly from Eq.~(\ref{eq:t-matrix_pg}) that the
dispersion of non-condensed pairs is of the form
\begin{equation}
 \Omega_q = \hbar^2 q^2 /2M_0^* \,.
\label{eq:mass}
\end{equation}
In summary, this quadratic 
dispersion can be derived from the pair susceptibility
$\chi(Q)$.  In turn, the particular form for $\chi(Q)$ shown in
Eq.~(\ref{eq:chi}) is chosen in
order to be consistent with Eqs.~(\ref{eq:gap_equation}) and
(\ref{eq:number_equation}). In this sense the usual
BEC constraint expressed by Eq.~(\ref{eq:chempot})
is intimately connected to the BCS-like gap equation
of Eq.~(\ref{eq:gap_equation}). The details
of this analysis are shown in Appendix
\ref{app:1}.

\subsection{Differences Between One and Two Channel Models}
\label{sec:2B}

We now extend this analysis
to include Feshbach bosons \cite{Griffin,Milstein}.
For this situation we can write down an equation  \cite{JS2} equivalent
to Eq.~(\ref{eq:gap_equation})
%(\ref{eq:gap_equation1})
with the direct fermion interaction $U$ replaced by $U_{eff} \equiv
U+g^2/(2 \mu-\nu)$ and effective scattering length $ a_s \rightarrow
a_s^*= mU^*/(4 \pi \hbar^2)$. Here we define
\begin{equation}
U^* \equiv U_0 - \frac{g_0^2}{(\nu_0 - 2 \mu)} \equiv \frac{4 \pi\hbar^2
  a_s^*}{m} \,,
\label{eq:C7}
\end{equation}
where $a_s^*$ is dependent on $\mu$.
We thus have 
\begin{eqnarray}
\Delta(T)& =&-U_{eff} \sum_{\bf k} \Delta(T)
 \frac{1-2 f(\Ek)}{2 \Ek}\,, \label{eq:18}\\
n & =& \sum _{\bf k} \left[ 1 -\frac{\ek - \mu}{\Ek}
+2\frac{\ek - \mu}{\Ek}f(\Ek)  \right] \,,
\label{eq:19}
\end{eqnarray}
where $\Ek = \sqrt{ (\ek -\mu)^2 + \Delta^2 (T) }$, and
again $\Delta$ is to
be distinguished from the order parameter \cite{JS2}.
Alternatively one can rewrite Eq. (\ref{eq:18}) as
\begin{equation}
\frac{m}{4 \pi \hbar^2 a^*_s } =  \mathop{\sum_{\bf k}}\left[ \frac{1}{2
\ek } - \frac{1- 2 f(\Ek)}{2E_{\bf k}} \right] \,.
\label{eq:18a}
\end{equation}

In Eq.~(\ref{eq:19}) $n$ represents the number of fermions, but the number
equation for the total number of particles 
involves both condensed and uncondensed bosons as well 
\begin{equation}
n + 2 n_b + 2 n^0_b = n^{tot} \;,
\label{number_equation}
\end{equation}
where $n^0_b=\phi _m^2$ is the number of molecular bosons in the
condensate.
The number of noncondensed molecular bosons is given by
\begin{equation}
n_b(T) = -\sum_{Q\ne 0} D(Q)
\label{eq:62a}
\end{equation}
where the Bose propagator is
\begin{equation}
D(Q) \equiv  \frac{1}{i\Omega_n - \epsilon_q^{mb} - \nu + 2 \mu  - \Sigma_B(Q)}.
\end{equation}
and we choose the self energy \cite{JS2}
\begin{equation}
 \Sigma_B (Q) \equiv -g^2 \chi(Q) / [ 1 + U \chi(Q) ]
 \label{eq:sigmaB}
 \end{equation}
to be consistent with the Hugenholtz-Pines condition that
bosons in equilibrium with a condensate must necessarily have zero
chemical potential.
This is equivalent to
\begin{equation}
\mu_{boson}(T) = 0,~~~~T \leq T_c\,,
\label{eq:chempot2}
\end{equation}
where $\mu_{boson} = 2 \mu - \nu + \Sigma_B(0)$. 
It follows after some simple
algebra that Eq.~(\ref{eq:chempot2}) is then consistent with
Eq.~(\ref{eq:18}).

\subsection{Parameter Choices and Renormalization Scheme}
\label{sec:2C}

We have introduced quantities ($U_0,g_0,\nu_0$) which characterize the
two body scattering in vacuum and correspond to the scattering lengths
$a_s$ or $a_s^*$.  How are these related \cite{Kokkelmans} to the
parameters ($U, g, \nu$) which enter into the Hamiltonian?  To begin we
ignore Feshbach effects.  The way in which the
%``bare"
two body interaction $U$ enters to characterize the scattering (in
vacuum) is different from the way in which it enters to characterize the
N-body processes leading to superfluidity.  In each case, however, one
uses a T-matrix formulation to sum an appropriately selected but
infinite series of terms in $U$.  For the two-body problem in vacuum,
there is a measurable characteristic of the scattering, the scattering
length, $a_s$,
\begin{equation}
\frac{m}{4 \pi\hbar^2 a_s} \equiv \frac {1}{U_0}
\label{eq:11g}
\end{equation}
which is related to $U$ via the Lippmann-Schwinger equation
\begin{equation}
\frac{m}{4 \pi\hbar^2 a_s } \equiv \frac{1}{U} + \sum_{\bf k} \frac{1}{2 \ek}
\label{eq:11c}
\end{equation}

In this way one may solve for the unknown $U$ in terms of $a_s$ or
$U_0$. The two-body T-matrix equation (Eq.(\ref{eq:11c})) can be
rewritten as
\begin{equation}
U = \Gamma U_0, ~~~~\Gamma = \left(1+ \frac {U_0}{U_c}\right)^{-1}
\label{eq:11h}
\end{equation}
where we define the quantity
$U_c$ as
\begin{equation}
\frac{1}{U_c} =- \sum_{\bf k} \frac{1}{2 \ek}
\label{eq:Uc}
\end{equation}
Here $U_c$ is the critical value of the potential
associated with the binding of a two particle state in vacuum.
Specific evaluation of $U_c$
requires that there be a cut-off imposed on the above
summation, associated with the range of the potential.

Now, for the Feshbach problem, the calculation proceeds in a similar
fashion and one finds \cite{Kokkelmans}
\begin{equation}
U=\Gamma U_0 , ~~~g = \Gamma g_0 , ~~~\nu - \nu_0 =  - \Gamma
\frac{g_0^2}{U_c} \,,
\label{eq:200}
\end{equation}
Here we define the parameter $\nu_0$ which is directly related to the
difference in the applied magnetic field $B$ and $B_0$
\begin{equation}
\nu_0 = (B-B_0) \Delta \mu^{0},
\label{eq:11x}
\end{equation}
where $\Delta \mu^{0}$ is the difference in the magnetic moment of the
two paired hyperfine states.  To connect the various energy scales,
which appear in the problem, typically $1 \textrm{Gauss} \approx 60 E_F$ for
$^{40}$K.

\subsection{Characterizing the Condensate}
\label{sec:2D}

There is an important difference in the nature of the condensate for the
two cases, with and without Feshbach bosons.  In the latter case there
are two components to the condensate associated with Cooper pairs
($\Delta_{sc}$) and condensed molecular bosons.  Moreover, the Cooper
condensate enters only into the gap equation, whereas the molecular Bose
condensate enters also into the number equation.  As long as the bosonic
condensate $n_b^0$ is non-negligible, this difference leads to essentially
different physics between the one and two channel problem, as will be
demonstrated below.
%Moreover, had one chosen to
%describe Feshbach effects by including two species of fermions
%one of which is taken to correspond to the Feshbach boson,
%the behavior is essentially that of a one channel system.

In this section we enumerate the different relationships between
the Cooper condensate and bosonic condensate terms.
The order parameter associated with Eq.~(\ref{eq:0c})
represents a linear combination of both paired fermions
(Cooper condensate) and
condensed molecules. It is given by \cite{Milstein,Griffin}
\begin{equation}
\tilde{\Delta}_{sc}=\Delta _{sc}-g \phi _m
\label{eq:C4}
\end{equation}
where, the boson order parameter $\phi _m=\langle b_{{\bf
 q}=0} \rangle$.
 We have
 \begin{equation}
 \Delta _{sc}=-U \sum _{\bf k}\langle a_{-{\bf k}\downarrow}a_{{\bf
  k}\uparrow}\rangle \,,
  \label{eq:C5}
  \end{equation}
therefore, $\tilde{\Delta}_{sc}$ can be written as
  \begin{equation}
  \tilde{\Delta}_{sc} =- \left[ U + \frac{g^2}{(2 \mu - \nu)} \right]
  \sum _{\bf k}  
  \langle a_{-{\bf k}\downarrow}a_{{\bf
   k}\uparrow}\rangle
   \label{eq:C6}
   \end{equation}
   where we have used the fact \cite{Kokkelmans} that
\begin{equation}
\phi _m=\frac{g \Delta _{sc}}{(\nu -2\mu) U} \,.
\label{eq:extra}
\end{equation}

The number of condensed Feshbach bosons which enters the
number equation [Eq.~(\ref{number_equation})] is given by
$n^0_b=\phi _m^2$.
Thus we have
\begin{equation}
n_b^0 = \frac{g^2 \Delta_{sc}^2}{[(\nu -2 \mu)U]^2}=
\frac{g^2 \tilde{\Delta}_{sc}^2} {[(2\mu-\nu)U+g^2]^2}
\label{eq:C2}
\end{equation}
Using Eqs.~(\ref{eq:200}) we can also write
\begin{equation}
n_b^0 = \left(1-\frac{U_0}{U^*}\right)^2
\frac{\tilde{\Delta}_{sc}^2}{g_0^2} \,,
\label{eq:C3}
\end{equation}
where $U^*$ is defined in Eq.~(\ref{eq:C7}).

\section{Equations of State at $T=0$}
\label{sec:3} 

\subsection{One Channel Model}
\label{sec:3A}

We now rewrite our central equations (\ref{eq:gap_equation}),
(\ref{eq:number_equation}) and (\ref{eq:14}) in the near-BEC limit to
compare more directly with the case of a weakly interacting Bose gas,
described by Gross-Pitaevskii (GP) theory.
%Our starting points should be contrasted
%however. Central to the present formalism is the behavior of
%the fermionic excitation gap $\Delta$,  as constrained
%by Eq.(\ref{eq:gap_equation}). This parameter is irrelevant for the Bose gas.
%Moreover, the primary interactions of the present approach 
%are inter-fermionic.  There are no direct boson-boson interactions,
%as in GP theory.  Finally, the ``boson"-like propagator of
%Eq.\ref{eq:t-matrix_pg}) is fundamentally dependent on the underlying fermionic
%character, and, notably, has a $q^2$ dispersion.
%This dispersion is intimately connected to the physics of
%the BCS gap equation.
%
It can be shown that (in the absence of FB)
\begin{equation}
n =  \Delta^2 \frac{  m^2} {4 \pi \sqrt{2m |\mu|} \hbar^3},
\label{eq:7}
\end{equation}
which, in conjunction with the expansion of Eq.~(\ref{eq:gap_equation}),
%(expanded in
%powers of $\Delta^2/\mu^2$) :
\begin{equation}
\frac{m}{4 \pi \hbar^2 a_s}=\left(\frac{2 m}{\hbar^2}\right)^{3/2}
\frac{\sqrt{|\mu|}}{8 \pi} 
\left[1+\frac{1}{16} \frac{\Delta^2}{\mu^2}\right],
\label{eq:gap_eq_exp}
\end{equation}
yields
%\begin{equation}
%n = \frac{\mu + \frac{\hbar^2}{2 m a_s^2}}{a_s \pi \hbar^2 /m}
%\end{equation}
%
\begin{equation}
\mu = - \frac{\hbar^2}{2 m a_s^2} + \frac{a_s \pi n \hbar^2} {m}.
\label{eq:mu_Uonly}
\end{equation}
These equations hold at all $ T \leq T_c$.
At $T=0$, these equations have been shown  \cite{Stringari,Strinati} to be
equivalent to the results of GP theory where one
identifies an effective inter-pair scattering length $a_B = 2a_s$
via $n_B = \frac{\mu_B }{4 \pi a_B \hbar^2 /M_B}$.
Here $n_B = n/2$ represents the number density of pairs, $\mu_B = 2\mu +
\hbar^2/ma_s^2$ is the ``bare" chemical potential of the pairs, 
and $M_B\approx 2m$
the pair mass.

\subsection{Effects of Feshbach bosons}
\label{sec:3B}

We now show that in the presence of Feshbach bosons this equation of
state is no longer that of GP theory and, moreover, there are important
differences in the ratio of the bosonic to fermionic scattering lengths.
As a result of the Bose condensate $n_b^0$ in
Eq.~(\ref{number_equation}) one finds an extra term in the number equation
\begin{eqnarray}
n^{tot} 
&=&\Delta^2 \left[\frac{  m^2} {4 \pi \sqrt{2m |\mu|} \hbar^3}+2
  \frac{(1-U_0/U^*)^2}{g_0^2} \right] .
\label{eq:density}
\end{eqnarray} 
%where we define $U^*=U_0+\frac{g_0^2}{2 \mu-\nu_0}$.
Combining the gap and number equation yields 
\begin{eqnarray}
\lefteqn{\frac{m}{4 \pi \hbar^2 a_s^*}=\left(\frac{2
      m}{\hbar^2}\right)^{\frac{3}{2}} 
\frac{\sqrt{|\mu|}}{8 \pi} }&& \nonumber \\ 
&&{}\times\left[1+\frac{1}{16 \mu^2}\frac{n^{tot}}{\frac{1}{16 \pi \sqrt{|\mu|}}
(\frac{2 m}{\hbar^2})^{\frac{3}{2}}+2 \frac{(1-U_0/U^*)^2}{g_0^2}}\right].   
\label{eq:combined}
\end{eqnarray}
Solving for $\mu$ in terms of $a_s^*$ one finds a new equation of state 
\begin{equation}
\mu \approx - \frac{\hbar^2}{2 m a_s^{*2}} + \frac{2 \pi^2}{m} \frac{g_0^2
  \hbar^2}{U_0^2} n^{tot} a_s^{*4} 
\end{equation}
to lowest order in $a_s^*$.  The second term in the above equation
derives from the Bose condensate term.  The first term in Eq.
(\ref{eq:density}) contributes a term of order $a_s^{*7}$ to this
correction \cite{Words1}.  This behavior should be contrasted with the
situation when FB are absent, where $a_B=2a_s$. It should also be
emphasized that the fermionic scattering length in a model without FB is
an independent experimental parameter, while here $a_s^*$ depends on
$\mu$, and must be obtained self-consistently.

\section{Calculations at $T_c$}
\label{sec:4}

We turn next to a calculation of $T_c$, which requires that we determine
$\Omega_q$ [via the T-matrix of Eq.~(\ref{eq:t-matrix_pg})] as a
function of the scattering length $a_s$. We address the one-channel case
first. The general expression for $1/M_0^*$ in the near BEC limit is
given by
\begin{equation}
\frac{1}{M_0^*}=\frac{1}{Z_0 \Delta^2} \sum_{\bf k}
\left[ \frac{1}{ m} \vk^2
- \frac{4 \Ek \hbar^2 k^2}{3m^2 \Delta^2} \vk^4\right] \,,
\label{eq:B}
\end{equation}
where we have used Eqs.~(\ref{eq:mass}) and (\ref{eq:expandt}).
After expanding to lowest order in $na_s^3$ 
\begin{equation}
M_0^* \approx 2m \left( 1 + \frac{\pi a_s^3 n}{2}\right) \,.
\label{eq:mass_change}
\end{equation}
From Eq.~(\ref{eq:14}) we then conclude that, at $T_c$,
\begin{equation}
\frac{n}{2}=\sum_{\bf q} b(\Omega_{\bf q}, T_c) \,.
\label{eq:totaln}
\end{equation}
Equation (\ref{eq:totaln}) reflects the fact that,
in the near-BEC limit, and at $T_c$, all fermions
are constituents of uncondensed pairs.  From the
above equation it follows that $(M_0^* T_c)^{3/2} \propto n = const.$
which, in conjunction with Eq.~(\ref{eq:mass_change}) implies
\begin{equation}
\frac{ T_c - T_c^0}{T_c^0} = - \frac{\pi a_s^3 n}{2}.
\end{equation}
Here $T_c^0$ is the transition temperature of the ideal Bose gas with
$M_0=2m$. This downward shift of $T_c$ follows the effective mass
renormalization, much as expected in a Hartree treatment of GP theory at
$T_c$. Here, however, in contrast to GP theory for a homogeneous
system with a contact
potential  \cite{RMP}, there is a non-vanishing renormalization of the
effective mass.

We turn now to the analysis of the behavior of $T_c$ in the presence
of Feshbach bosons. The (inverse) residue in the $T$-matrix is replaced by
\begin{equation} 
Z_1=Z_0+\frac{g^2}{\left[U(2 \mu-\nu)+g^2 \right]^2} \approx
\frac{n}{2\Delta^2} 
\end{equation}
and the $\hbar^2 q^2$ coefficient $B_1\equiv 1/(2 M_1^*)$ in $\Omegaq$ 
is such that 
\begin{equation}
B_1=\frac{B_0Z_0+\frac{1}{2 M}  \frac{g^2}{\left[U(2
      \mu-\nu)+g^2\right]^2}}{Z_1} 
%\approx \frac{1}{4m}\left[1-\frac{g_0^2}{U_0^2}\pi^2 n a^{*6}\right].
\approx \frac{1}{4m}\left[1-\frac{2g_0^4}{U_0^4}\pi^3 n a_s^{*9}\right].
\end{equation}
Here $Z_0$ and $B_0$ are the appropriate counterparts when FB are
absent.  Since $Z_0$ is proportional to the fermionic contribution to
the density, it is very small in the BEC limit.  Using the same
reasoning as in the previous case, we conclude that the ratio $T_c/B_1$
is constant with varying coupling.  \textit{Thus, again, $T_c$ follows
  the behavior of the inverse effective mass} with,
  to leading order, $(T_c^0-T_c)/T_c^0
\propto a_s^{*9}$.

\section{Particle Density Profiles in Traps}
\label{sec:5}

The differences between the equations of state for the one and two
channel models will have physical implications in the density profiles
of particles in a trap.  We now introduce the harmonic trapping
potential $V(r)=\frac{1}{2}m \omega^2 r^2$, which is treated in the
Thomas-Fermi (TF) approximation.  In this approximation, one replaces
$\mu$ with $\mu(r)=\mu-V(r)$.  In contrast to the uniform case, here
$\mu_{pair}(r, T)$ becomes non-zero beyond a critical radius $R_c(T)$,
where $R_c(T_c) =0$.  In this way at $T_c$, only the center of the trap
is superfluid, while at $T=0$ all of the trap contains condensed states.
To obtain the $T=0$ density profile, $n(r)$, we insert $\mu(r)$ into
Eq.~(\ref{eq:combined}) and solve for $n(r)$ [Here $n(r)$ refers to the
sum of both fermion and molecular boson contributions].  The solution is
\begin{eqnarray}
n(r)&=&16 \mu^2(r) \left[\frac{\hbar}{\sqrt{2m|\mu(r)|} a_s^*} -1\right]
\nonumber 
\\  &&\times{}   \left[\frac{1}{16 \pi \sqrt{|\mu(r)|}}(\frac{2
    m}{\hbar^2})^{\frac{3}{2}}+2 \frac{(1-\frac{U_0}{U^*})^2}{g_0^2}
    \right] ,
%n(r)=16 \mu^2(r) \left[\frac{m}{4 \pi \hbar^2 a^*}(\frac{\hbar^2}{2
%    m})^{\frac{3}{2}} \frac{8 \pi}{\sqrt{|\mu(r)|}}-1\right] \nonumber
%\\ 
%\cdot   \left[\frac{1}{16 \pi \sqrt{|\mu(r)|}}(\frac{2
%    m}{\hbar^2})^{\frac{3}{2}}+2 \frac{(1-\frac{U_0}{U^*})^2}{g_0^2}
%    \right] 
\label{eq:nr_FB}
\end{eqnarray} 
where we use $N=\int n(r)d^3r$ to self-consistently determine $\mu$.
Here it should be noted that $a_s^*$ is itself a function of $\mu(r)$.
%
%Without FB, Eq.~(\ref{eq:nr_FB}) becomes 
%
%\begin{equation}
%n(r) = \frac{1}{\pi}\left(\frac{2m|\mu(r)|}{\hbar^2}\right)^{\frac{3}{2}}
%\left[\frac{\hbar}{\sqrt{2m|\mu(r)|} a} -1\right].
%\label{eq:nr_nonFB}
%\end{equation}
% 
%It should be noted that
%at finite $T$, the gap equation (\ref{eq:gap_equation}) is valid only
%for $r<r_c$, where the critical radius $r_c$ is $T$ dependent.

\begin{figure}
\centerline{\includegraphics[width=3.0in,clip]{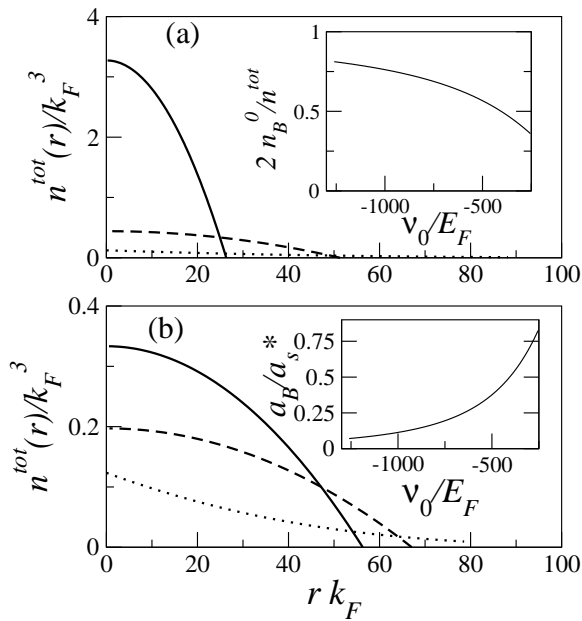}}
\caption{
  Density profiles at (a) $\nu_0 = -1260$ and (b) $\nu_0 = -250$.  Solid
  and dashed lines are at $T=0$ with and without Feshbach bosons (FB),
  respectively, and the dotted lines are at $T=T_c$.  Inset to (a) plots
  the FB weight in the condensate.  Inset to (b) shows the ratio of the
  inter-boson to inter-fermion scattering lengths at trap center.  Units
  are chosen such that 
%$\hbar=1$, $2m=1$, 
  $k_F\equiv 1$, and $E_F=(3 N)^{\frac{1}{3}} \hbar \omega=1$. Hence the units
  for $U_0$, $E_F/k_F^3$, and for $g_0$, $E_F/k_F^{3/2}$, are both set to
  unity.  For $\nu_0=-250$ the interaction parameter $k_F a_s^* \approx
  0.22$.}

\label{fig:profiles}
\end{figure}

In Fig.~\ref{fig:profiles}(a), we plot $n(r)$ for the parameters
$U_0=-0.89$, $g_0=-35$, $\nu_0=-1260$ and $N=10^5$.  For this value of
$\nu_0$ we are somewhat away from the deepest BEC regime  as is necessary 
to ensure the validity of the TF approximation and the
fermionic contribution to the density is no longer negligible compared
to its bosonic counterpart. Indeed, the percentage weight of the
molecular boson condensate (shown in the inset) indicates that by $\nu_0
\approx -400$ the Cooper pair condensate is beginning to dominate.
%this is necessary to ensure the validity of the Thomas Fermi
%approximation, where the effective inter-bosonic scattering length,
%$a_B$ must satisfy $N a_B/a_{ho} >> 1 $, in order for the kinetic energy
%of the gas to be dominated by its potential energy.  Here
%$a_{ho}=\sqrt{\frac{\hbar}{m \omega}}$. 
In this figure we also show the density profile as computed in the
absence of FB (dashed line), as well as the behavior at $T=T_c$ (dotted
line).  Our $T=T_c$ curves were computed in the absence of FB for both
panels. We were unable, thus far to find important differences between
this distribution and that of an ideal Bose gas. Here one has to solve
self consistently for $\mu_{pair}(r)$, as well.
%\begin{figure}
%\centerline{\includegraphics[angle=0,width=3.0in]{ratio.eps}}
%\caption{Ratio of bosonic to fermionic scattering lengths.}
%\label{fig:ratio}
%\end{figure}
%

To arrive at a meaningful comparison of the two $T=0$ cases (with and
without Feshbach bosons) we used the same value for the effective
two-body scattering length $a_s^*$, obtained from the self-consistent
calculations of $n(r)$ in the presence of FB. The profile without FB
bosons is then calculated using the familiar TF result near the BEC
limit  \cite{Stringari} at a fixed fermionic scattering \cite{Words2}
length $a_s^*$.
%= a$
%m (U_0 +\frac{g_0^2}{2 \mu - \nu_0})/(4\pi \hbar^2)$.  
The same plots are presented with $\nu_0 = -250$ in
Fig.~\ref{fig:profiles}(b), which is further from the BEC limit.
%It may be seen from both panels that in the presence of FB, the density
%profiles are narrower.
  
An important consequence of the one \textit{vs} two channel problems is
that for the latter, as the extreme BEC limit is approached, the number
of fermions is diminished in favor of FB. This occurs due to the self
consistent adjustment of the fermionic chemical potential.  Thus the
interaction between the bosons mediated by the fermions is weaker than
what might have been expected without consideration of FB.  This
weakened interaction is reflected
% \cite{RMP} 
in Fig.~\ref{fig:profiles} through a comparison
between the solid and dashed lines, which shows that the trap profiles
are narrower in the presence of Feshbach effects.  Indeed, this could
have been anticipated from the above calculations in the
homogeneous case, at $T=0$ and $T_c$,
which deduced a very weak dependence of the bosonic scattering length
and effective pair mass on the fermionic scattering length $a_s^*$.
%CHECK
This finding mirrors the result in a trapped atomic Bose
gas\cite{RMP} where one sees
that a weaker repulsive inter-boson interaction leads to
a narrowed density profile.
Comparison between the lower and upper panels of Fig.~\ref{fig:profiles}
shows that in both cases (with and withour FB), the profiles become
narrower as the BEC limit is approached.

We may use the results in Fig.~\ref{fig:profiles} to obtain a
semi-quantitative estimate of the bosonic scattering length $a_B$, based
on a phenomenology used in experimental
analysis \cite{Jin3,Grimm,Ketterle2}.  We compare our results to the
Thomas-Fermi approximated GP equation at $T=0$ which yields
%\begin{equation}
%(V(r)+ U_B\Psi^{\dagger}(r)\Psi (r))\Psi(r)=\mu_B \Psi(r)
%\end{equation}
$n_B(r)=\frac{\mu_B-V(r)}{U_B}$ where the inter-boson interaction $U_B$
is connected to the scattering length $a_B$ via $U_B=\frac{4 \pi\hbar^2
  a_B}{M_B}$.  If we fit the profiles of Fig.~\ref{fig:profiles} to the
inverted parabola ($C_1-C_2r^2$), we may infer that $a_B=\frac{M_B^2
  \omega^2}{4 \pi \hbar^2 C_2}$.  The ratio $a_B/a_s^*$ is plotted vs
$\nu_0$ in the inset to the lower panel of Fig.~\ref{fig:profiles}b.
The same analysis applied to the profile without FB yields the familiar
2:1 ratio of the bosonic to fermionic scattering lengths.  The parameter
$a_B$ is an important quantity which appears in experiment to be
considerably less than a factor of 2 times its fermionic counterpart.

\section{Conclusions}
\label{sec:6}
There are very detailed calculations \cite{Petrov} of four-body atomic
scattering processes which yield a fixed length ratio of $0.6$ for the
bosonic to fermionic scattering lengths.  Presumably in the asymptotic
BEC regime this constraint should be imposed on a more exact theory of
the ground state. Indeed, a weakness of the mean field approach taken
here (based on a generalized BCS ground state) is that the
``inter-boson" scattering length is treated only approximately. This can
be viewed as related to the inclusion of only one fermion and two
fermion propagators
(\textit{i.e}, T-matrix) with no higher order terms.

A strength of this approach, over the counterpart atomic calculations is
that many body physics associated with the broken symmetry or superfluid
state is included.  \textit{A comparison of the one and two channel calculations
presented here shows that, through many body effects, the nature of the
condensate enters in an important way to determine the equation of state
and, thereby, the ratio of the scattering lengths}.  It should also be
stressed that the many body context with which the number $0.6$ is to be
compared is different from $2.0$, which is the number generally assumed
in the literature. This difference appears once one includes Feshbach
bosons.  Indeed, while it depends upon the magnetic field or detuning,
in the near-BEC regime this ratio is found to be significantly less than
$2.0$. This is a central point of the present paper.  Future experiments
will be required to determine, whether, as some have presumed, the ratio
of $0.6$ applies to all detunings, or whether
this number is variable as argued
here.

In this paper we have shown that superfluidity of fermionic atoms in the
near-BEC limit is in general different from Bose superfluidity (as
described by GP theory). We have compared one and two channel models and
find that differences from the GP picture for each are associated with
the underlying fermionic character of the system. This shows up at $T
\neq 0$ for the one channel system and at all $T$ for a Hamiltonian
which includes Feshbach bosons.

Essential to our approach is the presence (at all $T \ne 0$) of
non-condensed pairs which contribute separately to the fermionic gap
parameter $\Delta$, in addition to the condensate term.  Noncondensed pairs or
``pseudogap effects" have been generally ignored in the
literature \cite{NSR} and we have shown elsewhere \cite{JS2} that they are
negligible only in the strict BCS regime. Our inclusion of these effects
corresponds  \cite{JS} to a self-consistent Hartree treatment of pairing
fluctuations in BCS theory, in which condensation naturally occurs in
the presence of a finite excitation gap (``pre-formed pairs").  Our
approach also makes evident the strong analogies between this
generalization of BCS theory and Bose condensation, where both condensed
and non-condensed bosons must be properly characterized.  \textit{In the
  two channel model, non-condensed bosons should be viewed as consisting
  of a strongly hybridized mixture of fermion pairs and Feshbach bosons.
  In the one channel model they result exclusively from fermion pairs}.

A key aspect of this work is in the comparison we have presented between
one and two channel models.  Differences arise because of the nature of
the condensate. A fermionic or Cooper pair condensate ($\Delta_{sc}$)
enters into the gap equation while a bosonic condensate ($n_b^0$) enters
also into the number equation. When there is an appreciable fraction of
bosonic condensate (as in the near-BEC and BEC regimes) these
differences will be apparent. A distinction between the one and two
channel models is relatively unimportant once the bosonic condensate is
negligible (usually when the scattering length is large and positive or
alternatively negative). These differences are associated with physical
properties such as the equations of state and the density profiles in a
trap.  Feshbach bosons can, in effect, collapse much more completely to
the center of a trap than can fermion pairs.  This effect can be
inferred from the ratio of the scattering lengths, and is related to the
fact that in this limit, fermions are essentially absent, since the
number equation constraint can be entirely satisfied by populating the
bosonic state. As a result, these bosons are close to ideal.

For the relatively broad Feshbach resonances currently under study in
lithium and potassium, differences between the one and two channel
models are presumably only important in the BEC and near-BEC limits
studied here, where there is an appreciable bosonic condensate.  It will
be interesting to study narrower Feshbach resonances where these
differences may persist into the unitary scattering regime.

\acknowledgments

We acknowledge useful discussions with R. Hulet, D. Jin and Z.
Tesanovic.  This work was supported by NSF-MRSEC Grant No.~DMR-0213765
(JS and KL), NSF Grant No.~DMR0094981 and JHU-TIPAC (QC).

\appendix

\section{Connection between $\chi(Q)$ and the standard Gap Equation}
\label{app:1}
The expansion of the T-matrix at small four-vector $Q = (\Omega, {\bf
  q})$ can be written in the following form:
\begin{equation}
t^{-1}(Q)=Z_0(\Omega-\Omega_{\bf q}+\mu_{pair}+i \Gamma_{\bf q})
\end{equation}
Therefore, the condition for the divergence of the T-matrix
at zero $Q$ is equivalent to
\begin{equation}
\mu_{pair}=0
\label{eq:A1}
\end{equation}
or
\begin{equation}
U^{-1}+\chi(0)=0.
\label{eq:sfl_condition}
\end{equation}
We now show that for a proper choice of the pair susceptibility 
\begin{equation}
\chi(Q)=\sum_K G(K)G_0(Q-K) 
\label{eq:A2}
\end{equation}
Eq.~(\ref{eq:A1}) is equivalent to the BCS-like gap equation
of Eq.~(\ref{eq:gap_equation}),
provided also the Greens function is of the BCS form
\begin{equation}
G(K)=\frac{\uk^2}{i \omega_n-\Ek}+\frac{\vk^2}{i \omega_n+\Ek} =
\frac{i\omega_n + \xi_{\bf k}}{(i\omega_n)^2-\Ek^2} \,.
\end{equation}
Just as in BCS theory, we associate this dressed 
Green's function with self-energy 
\begin{equation}
\Sigma(K)=-\Delta^2 G_0(-K) 
\label{eq:A3}
\end{equation}

We calculate $\chi(Q=(i \Omega_m,{\bf q}))$ by 
performing the appropriate Matsubara sums following standard procedure
 \cite{Fetter}. At $Q=0$, we have 
%
%\begin{widetext}
\begin{eqnarray}
\chi(0)&\equiv& \sum_K \frac{i\omega_n + \xi_{\bf k}}{(i\omega_n)^2-\Ek^2}
\frac{1}{-i \omega_n-\xi_{\bf k}}  \nonumber\\  
%& = & -\sum_K \frac{1}{(i\omega_n)^2-\Ek^2} \nonumber\\ 
&=& \sum_{\bf k}\frac{1-2 f(\Ek)}{2 \Ek} 
%\chi(Q)&=&\sum_K  \left[\frac{\uk^2}{i \omega_n-\Ek}+\frac{\vk^2}{i \omega_n+\Ek}\right]
%\frac{1}{i \Omega_m-i \omega_n-\xi_{\bf k-q}} \nonumber \\
%%&=& \sum_K \left\{ \left[\frac{1}{i \omega_n-\Ek}-\frac{1}{i\omega_n-i\Omega_m+\xi_{\bf k-q}} \right]
%%\frac{\uk^2}{i\Omega_m-\Ek-\xi_{\bf k-q}}
%%+\left[ \frac{1}{i\omega_n+\Ek}-
%%\frac{1}{i \omega_n-i\Omega_m+\xi_{\bf k-q}}\right]\frac{\vk^2}{i \Omega_m+\Ek-\xi_{\bf k-q}}\right\}\nonumber \\ 
%&=&-\sum_{\bf k} \left[\uk^2 \frac{1-f(\Ek)-f(\xi_{\bf k-q})}{i \Omega_m-\Ek-\xi_{\bf k-q}}+\vk^2\frac{f(\Ek)-f(\xi_{\bf k-q})}
%{i\Omega_m+\Ek-\xi_{\bf k-q}}\right] 
\end{eqnarray}
%\end{widetext}
%
%Taking the $Q \rightarrow 0 $ limit, it follows that
%
%\begin{equation}
%\chi(0)=\sum_{\bf k}\frac{1-2 f(\Ek)}{2 \Ek}
%\end{equation}
and Eq.~(\ref{eq:sfl_condition}) becomes
\begin{equation}
U^{-1}+\sum_{\bf k}\frac{1-2 f(\Ek)}{2 \Ek}=0,
\end{equation}
which, in conjunction with the two-body scattering equation
\begin{equation}
\frac{m}{4 \pi \hbar^2 a_s}=U^{-1}+\sum \frac{1}{2 \ek},
\label{eq:A6}
\end{equation}
gives
\begin{equation}
\frac{m}{4 \pi \hbar^2 a_s}=\sum_{\bf k} \left[\frac{1}{2 \ek}-\frac{1-2 f(\Ek)}{2 \Ek} \right]
\label{eq:A5}
\end{equation}
Thus, we have demonstrated that, if we presume Eqs.~(\ref{eq:A2}) and
(\ref{eq:A3}), then Eqs.~(\ref{eq:A1}) and (\ref{eq:A5}) are equivalent.

%\input Appendix.tex

%\bibliographystyle{prsty}
%\bibliography{Review,Words}

\begin{thebibliography}{10}
\bibitem{Words3}
These Fermi functions can be set to zero, since their arguments are of the
  order of $|\mu|/T_c>>1$.
\bibitem{Words1}
If one expands the inverse $T$-matrix in the form of a GP equation, it can be
  seen that $a_B \propto \Delta^4 a_s^{*3}/n^2 \propto a_s^{*7}$.
\bibitem{Words2}
This scattering length should be distinguished from the isolated two-body
  scattering parameter %$ a = $m(U_0 -\frac{g_0^2}{ \nu_0})/(4 \pi \hbar^2)$
  which unlike $a_s^*$ does not reflect the interactions responsible for
  superfluidity. Indeed this isolated two-body parameter appears to be
  inappropriate, in that it becomes negative deeper into the BEC regime ($
  \nu_0 \leq \frac{g_0^2}{U_0}$). This sign change should occur quite generally
  % in any parameter set $(U_0 < 0)$, which as $U_0 < 0$ is required to obtain
  BCS-like behavior at large positive $\nu_0$.
\end{thebibliography}
%\begin{thebibliography}{}

%\end{thebibliography}

\end{document}